\documentclass[12pt]{article}   
\usepackage{psfig,times,fancyheadings}  

\begin{document}
\thispagestyle{empty} 

\baselineskip=20pt  
 \lhead[\fancyplain{}{\sl }]{\fancyplain{}{\sl }}
 \rhead[\fancyplain{}{\sl }]{\fancyplain{}{\sl }}

 \renewcommand{\topfraction}{.99}      
 \renewcommand{\bottomfraction}{.99} 
 \renewcommand{\textfraction}{.0}


\newcommand{\nc}{\newcommand}

\nc{\qI}[1]{\section{{#1}}}
\nc{\qA}[1]{\subsection{{#1}}}
\nc{\qun}[1]{\subsubsection{{#1}}}
\nc{\qa}[1]{\paragraph{{#1}}}

\def\qbu{\hfill \par \hskip 6mm $ \bullet $ \hskip 2mm}
\def\qee#1{\hfill \par \hskip 6mm #1 \hskip 2 mm}

\nc{\qfoot}[1]{\footnote{{#1}}}
\def\qL{\hfill \break}
\def\qpar{\vskip 2mm plus 0.2mm minus 0.2mm}
\def\qtvi{\vrule height 2pt depth 5pt width 0pt}
\def\qth{\vrule height 12pt depth 0pt width 0pt}
\def\qtb{\vrule height 0pt depth 5pt width 0pt}
\def\tvi{\vrule height 12pt depth 5pt width 0pt}

\def\qparr{ \vskip 1.0mm plus 0.2mm minus 0.2mm \hangindent=10mm
\hangafter=1}

\def\qdec#1{\par {\leftskip=2cm {#1} \par}}

\def\qdpt{\partial_t}
\def\qdpx{\partial_x}
\def\qddpt{\partial^{2}_{t^2}}
\def\qddpx{\partial^{2}_{x^2}}
\def\qn#1{\eqno \hbox{(#1)}}
\def\qds{\displaystyle}
\def\qw{\widetilde}
\def\qmax{\mathop{\rm Max}}   
\def\qmin{\mathop{\rm Min}}   


\def\qci#1{\parindent=0mm \par \small \parshape=1 1cm 15cm  #1 \par
               \normalsize}

\null
{\large \it Presented at the Third Nikkei Conference
on Econophysics, \qL Tokyo, November 10, 2004}
\vskip 0.6 cm

\centerline{\bf \Large Macro-players in stock markets}
\vskip 3mm

\vskip 0.5cm
\centerline{\bf Bertrand M. Roehner $ ^1 $ }
\vskip 3mm
         
\centerline{\bf Institute for Theoretical and High Energy Physics}
\centerline{\bf University Paris 7 }

\vskip 7mm

{\bf Abstract}\quad 
It is usually assumed that stock prices reflect a balance
between large numbers of small individual sellers and buyers. However,
over the past fifty years mutual funds and other institutional
shareholders have assumed an ever increasing part of stock
transactions: their assets, as a percentage of GDP, have been multiplied
by more than one hundred.
The paper presents evidence which shows that
reactions to major shocks are often dominated by a small
number of institutional players. Most often the market gets a wrong
perception and inadequate understanding of such events because
the relevant information (e.g. the fact that one mutual fund has 
sold several million shares)
only becomes available weeks or months after the event, through
reports to the Securities and Exchange Commission
(SEC). Our observations suggest that there is a radical difference
between small ($ < 0.5\% $) day-to-day price variations which may
be due to the interplay of 
many agents and large ($ >5\% $) price changes which, on the
contrary, may be caused by massive sales (or purchases) by a few
players. This suggests that the mechanisms which account for 
large returns are markedly different from those ruling small returns.

\vskip 0.3cm

\centerline{September 23, 2004}

\vskip 2mm
\centerline{\it Preliminary version, comments are welcome}

\vskip 0.3cm
Key-words: stock markets, mutual funds, investment funds, hedge-funds,
institutional owners, collective behavior, herd effect
\vskip 0.5cm 

1: ROEHNER@LPTHE.JUSSIEU.FR
\qL
\phantom{1: }Postal address where correspondence should be sent:
\qL
\phantom{1: }B. Roehner, LPTHE, University Paris 7, 2 place Jussieu, 
F-75005 Paris, France.
\qL
\phantom{1: }E-mail: roehner@lpthe.jussieu.fr
\qL
\phantom{1: }FAX: 33 1 44 27 79 90

\vfill \eject

\qI{Introduction}

Very broadly speaking, there are two ways to represent stock markets
and also two different methodologies to choose between them (Fig. 1).
In the micro-player representation, the number of players is large
enough to be treated by using statistical methods. In this case, each
individual player has only a negligible impact on daily price changes.
On the contrary, in the macro-player representation, the number of
players is small and each one has a substantial impact not only on daily
price changes but even on weekly or monthly price changes. In the second
case a game theoretic approach would be more sensible than
a statistical approach. 
The main objective of this paper is to find out which of these
descriptions corresponds to the situation of  markets in 2004. 
A first hint is provided
by the sheer weight of the macro-players. In 1900, the share of financial 
institutions in total corporate stock outstanding was 6.7\%, in 1974
it was 33\%, in 2002 it was of the order of 50\% (Kotz 1978, Statistical
Abstract of the United States 2003, p. 755). 
\qpar
  \begin{figure}[htb]
    \centerline{\psfig{width=12cm,figure=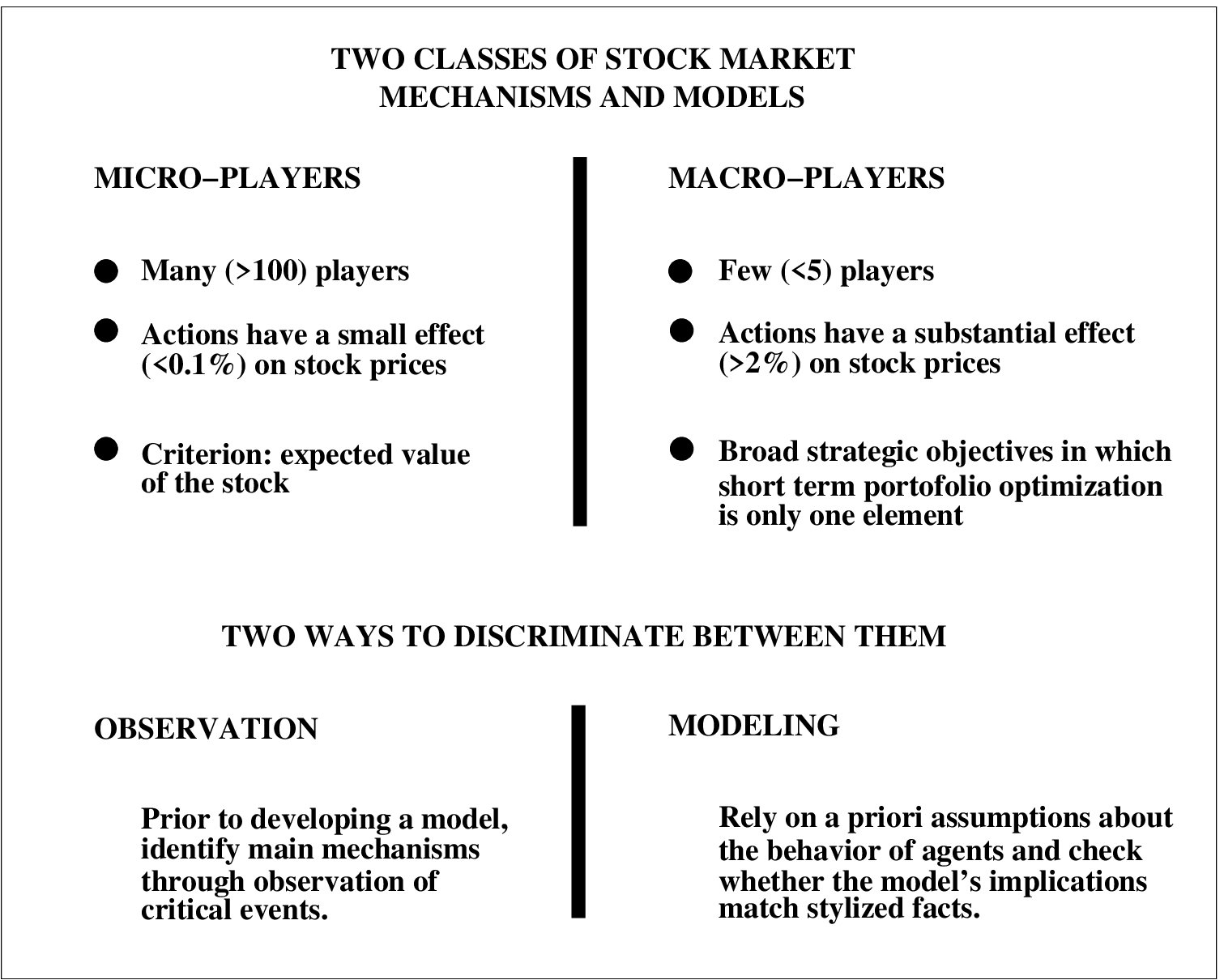}}
    {\bf Fig.1 }
{\small In this paper we want to discriminate between the micro- and
macro-player representations by observing the reactions of stocks to
major shocks. Trying to unravel market mechanisms 
prior to any attempt at constructing
specific mathematical models can be labeled as ex-ante analysis,
as opposed to ex-post analysis which in econometrics
is the standard approach.}
{\small \it }
 \end{figure}

The purpose of this paper is to show that many (though not all) 
important phenomena that occur nowadays in stock markets belong
to the second class. In order to make this point, we will use an approach
which can be labelled as an ex-ante analysis. In what sense does it differ from
the more commonly used ex-post analysis? In the ex-post analysis, one
begins by building a mathematical model whose predictions are then 
compared to a number of stylized facts. In econometrics, this is the 
approach which 
is used almost exclusively. In the ex-ante analysis one tries to
design an ``experiment'' whose results give us a better insight into
the mechanisms which are at work. When using the term ``experiment''
I do not mean a laboratory experiment with paid human subjects
but rather a problem-oriented observation 
(also referred to as a quasi-experiment) selected and designed in order
to shed new light on a specific phenomenon.
In this paper I will use the ex-ante analysis in order to decide which of
the micro- or macro-representation is more acceptable. More specifically,
I will emphasize the importance of strategic investments as opposed 
to transactions based on expected value. In addition,
by monitoring as closely as possible
the behavior of shareholders in the weeks and days preceding
a bankruptcy, I will analyze how investors
react to the risk of bankruptcy.
Although the paper relies on a number of case studies, I
believe that the behaviors which will be identified have a
fairly broad validity.
To begin with, I consider the case of Kmart,
the American retail store company.

\qI{Kmart: background information}

As several of the cases to be considered below concern Kmart, it is in order
to give some background information for this company. It was founded
in 1899 by Sebastian Kresge and was called the Kresge company until
1977 when its name was changed to Kmart. As shown by Fig. 2a it has
been a highly successful discount retailer for many decades, but
fell into trouble in the 1980s
\qfoot{Symbolizing this trend was the fact that
back in 1988, in the Oscar-winning film ``Rainman'', the character
played by Dustin Hoffman repeatedly refers to Kmart by saying 
``Kmart sucks'' meaning that the stores were shabby and displayed
low quality items.}
. 
Eventually it had to ask for Chapter 11 bankruptcy protection in January
2002. 
  \begin{figure}[htb]
    \centerline{\psfig{width=12cm,figure=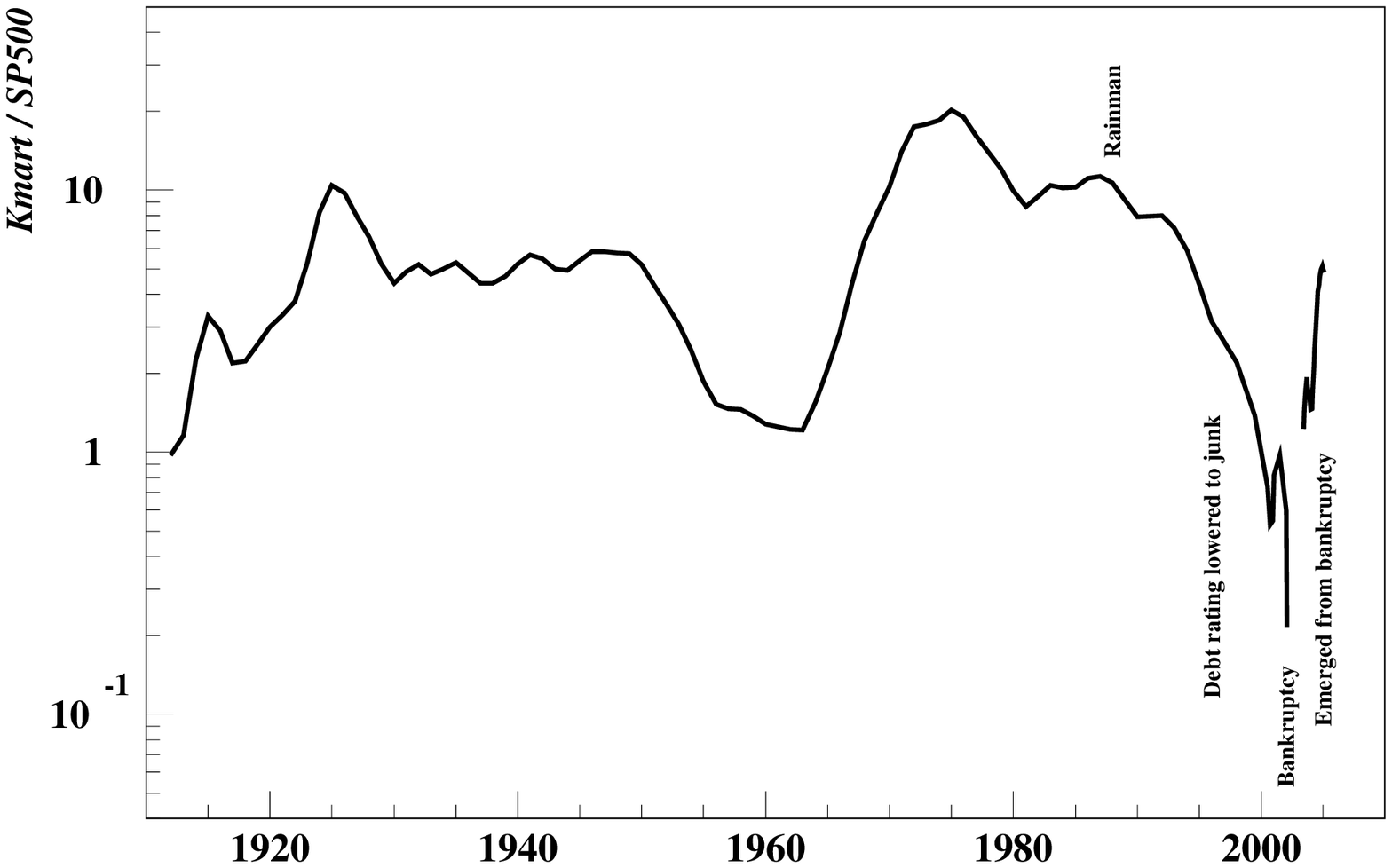}}
    {\bf Fig.2a Ratio of Kmart stock price to the Standard and Poor's
500 index (1910-1999)}
{\small In the 1950s In the late 1980s after becoming confronted to 
Wal-Mart's competition Kmart entered a downward spiral that lasted over
20 years and eventually lead to its bankruptcy in January 2002. 
After the company emerged from bankruptcy in May 2003, its shareprice
increased more than 5 times within 18 months.
The reference to the film ``Rainman'' is explained in the text.}  
{\small \it Source: Common stock (1992), Kmart Fact Book (1999)}.
 \end{figure}
The losing battle that Kmart fought against Wal-Mart
can be summarized by the following figures.

$$ \matrix{
\tvi 
 \hbox{}\hfill & \hbox{Kmart} \hfill & \hbox{Wal-Mart} \cr
\noalign{\hrule}
\qth 
\hbox{Revenue 1990 [billion dollars]} & 32 & 32 \cr
\hbox{Revenue 1994 [billion dollars]} & 36 & 83 \cr
} $$
\qpar

In 1993, Kmart had to close 5\% of its stores and in 1994 it experienced
a loss of one billion dollars. These poor performances led to increased
indebtedness and in 1996, the rating of its debt was lowered below
investment grade. For a company of the size of Kmart to be rated
at junk level is something which is not common. In subsequent years,
Kmart continued to lose market shares to Wal-Mart.
The fall of its share price shown in Fig. 2a is consistent with this loss of
momentum. However, the trajectory of the stock price shown in Fig. 2b,c 
is fairly puzzling and raises the following questions:
\qbu Why did it increase by almost 100\% between September 2000 and
August 2001? 
\qbu Why did it abruptly drop in January 2002 leading the company
into bankruptcy?
\qpar

In the expected value framework one would wish to know which innovations
in Kmart's growth perspectives justified these changes. In fact, there were
none. Both the increase and the sharp fall were due to causes which had
very little to do with Kmart's growth perspectives.
The 100\% rise resulted from the strategic move of a single investor,
Ronald Burkle, a billionaire and head of an investment firm.

\qI{Burkle's deal with Kmart}

Between October 2000 and October 2001, Burkle bought 7.2\% of Kmart's
outstanding shares (Fig. 2b).
  \begin{figure}[htb]
    \centerline{\psfig{width=12cm,figure=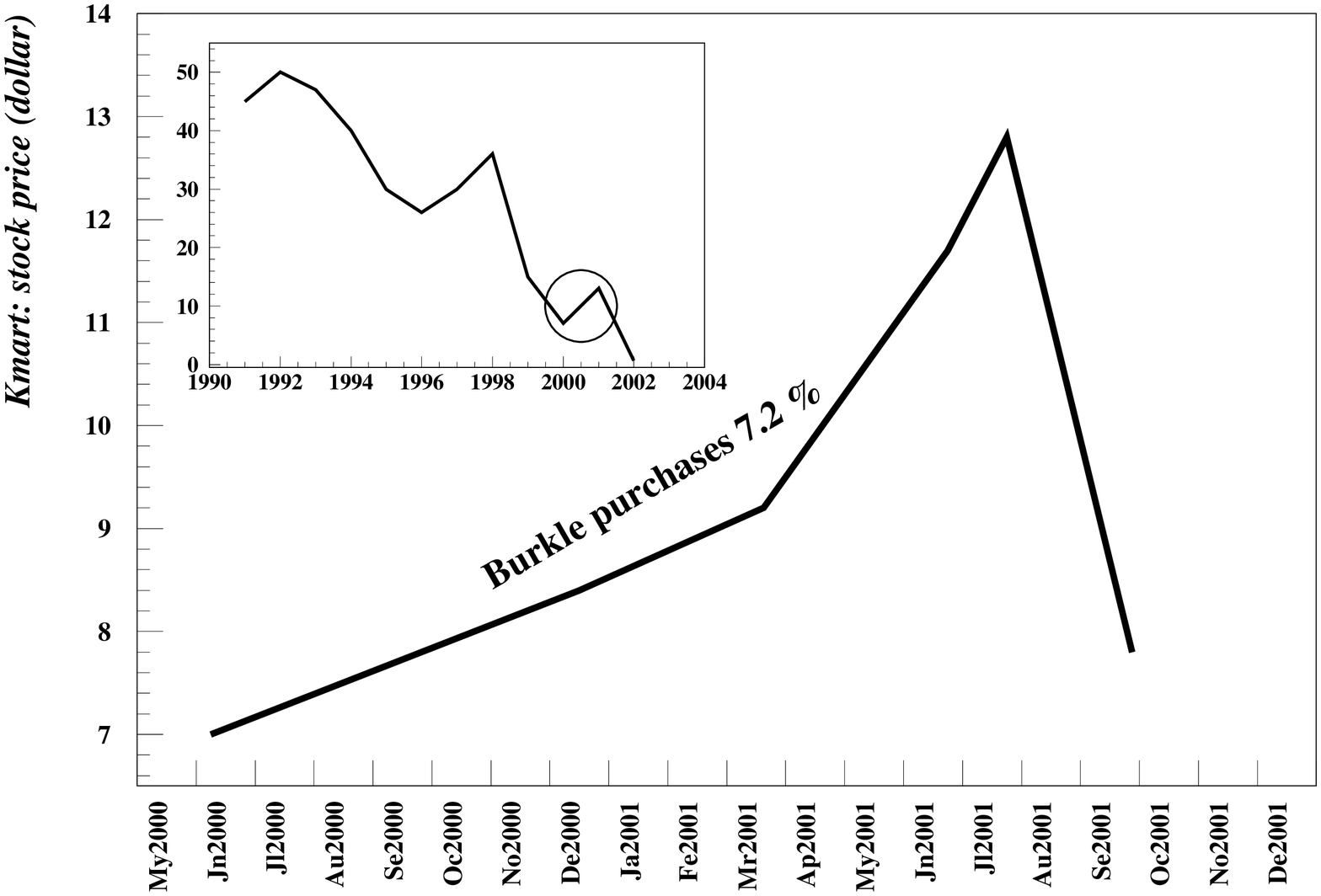}}
    {\bf Fig.2b Kmart share price (July 2000 - December 2001).}
{\small The insert shows how the price increase that took place 
between June 2000 and  August 2001 (as delimited by the circle)
fits into the broader picture; this increase was mainly due to the fact that
a consortium led by  billionaire Ronald Burkle purchased 7.2\%
of Kmart's outstanding shares. Once Burkle's share purchases stopped
the price resumed its downward trend.}  
{\small \it Source: CNN Money (January 22, 2002)}.
 \end{figure}
We know this because an investor who wishes
to buy more than 5\% of the shares has to notify the Securities and
Exchange Commission in advance which he did on October 13, 2000.
In the present case we are fortunate to know his global strategy, something
which is rarely the case. 
The purchase of Kmart shares was in fact
part of a broader deal. Before 2001, Kmart had two grocery suppliers:
Supervalue (for \$ 2.3 billions) and Fleming (for \$ 1.3 billion). In February
2001, that is to say 5 months after Burkle began his massive
purchases, it became known that Kmart had chosen Fleming as 
exclusive supplier for the next 10 years, a deal whose value
was estimated at \$45 billions. As it happens, Burkle had a stake of
almost 10\% in Fleming. This makes the deal fairly clear.
Burkle invests about \$ 0.1 billion in Kmart shares and in return he gets
an exclusivity contract that is worth 450 times more (CNN, February 7, 2001)
\qpar

Naturally, Kmart denied that there was any link between the two transactions.
However, one should keep in mind that in 1999 Kmart tried to 
initiate a buyback program of its own shares
for a total amount of \$ 1 billion;
assuming a price range from \$ 5 to \$ 10 per share, this
represented between 20\% and 40\% of its outstanding shares. 
Unable to complete 
this program by itself because of its indebtedness, Kmart certainly
relied on the deal with Burkle for implementing its objective, albeit
on a smaller scale than planned initially. 
\qpar

To sum up, the 100\% price increase in Fig.2b had much to do with
Burkle and Fleming, but very little with Kmart itself. We now turn to the
events which occurred in the weeks before Kmart's bankruptcy.

\qI{The withdrawal of Fidelity from Kmart}

In a CNN financial report of February 15, 2002 one reads: 
\qdec{Jim Lowell, editor of the newsletter ``Fidelity Investor'' said 
Fidelity recently slashed its holding in Kmart. Kmart had represented
almost 10\% of assets at Fidelity parent company Fidelity Management
and Research Corporation (FMR) until recently, before the company slashed
its position to 1.3\%.}
\qpar

We posit that Fidelity's move (which was certainly imitated by
other institutional holders even though we don't have explicit statements)
accounts for much of Kmart's stock price collapse in
January 2002 (Fig. 2c). Note that as holdings are reported only every 
three months we do not know when exactly the sales occurred. 
As a result the term
``recently'' used in the above excerpt of the CNN report
is fairly elastic: it refers
to a date comprised between November 15, 2001 and January 22, 2002.
  \begin{figure}[htb]
    \centerline{\psfig{width=12cm,figure=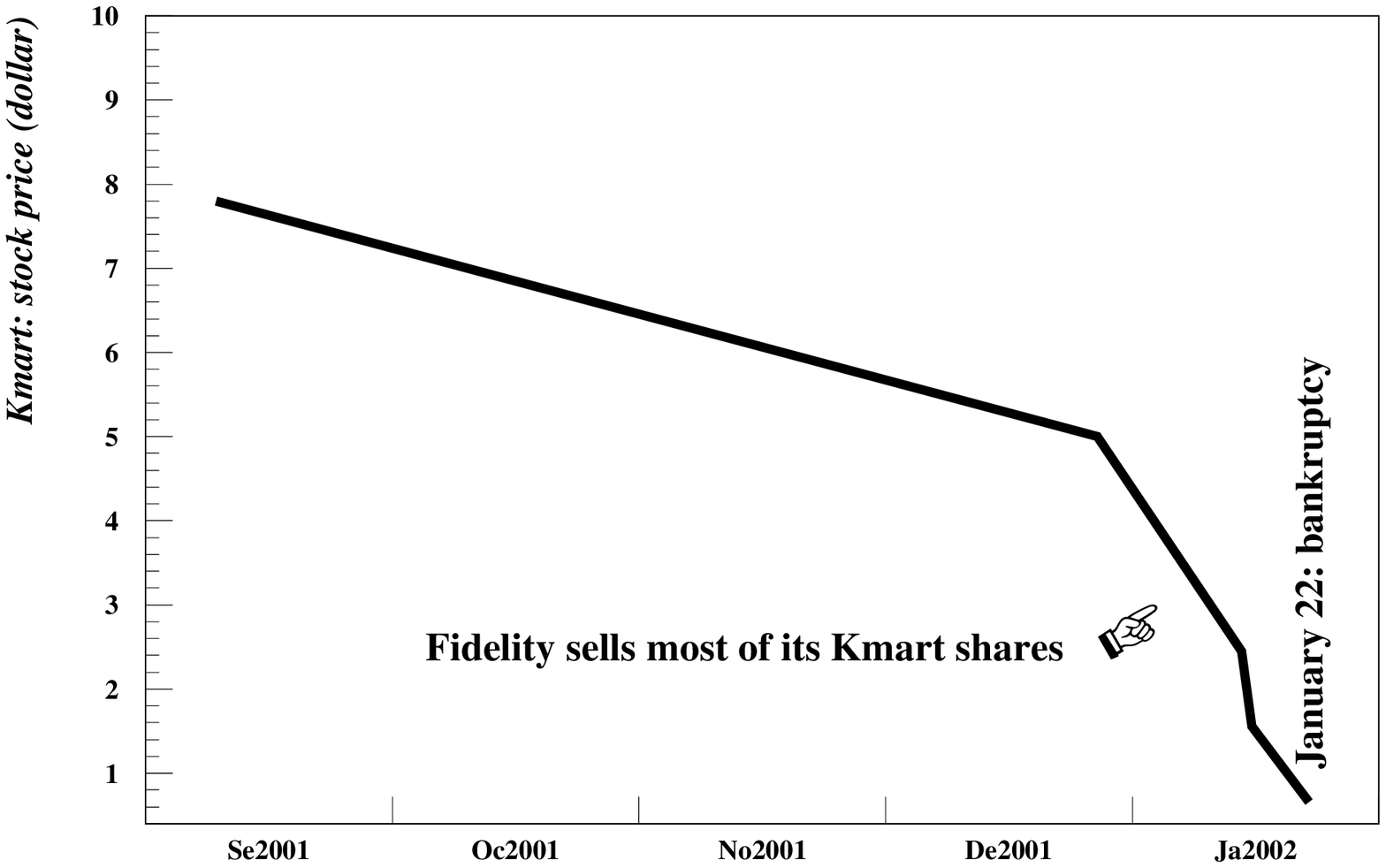}}
    {\bf Fig.2c  Kmart share price in January 2002.}
{\small When Fidelity began to sell its shares, it became obvious that
Kmart no longer had the support of its major institutional shareholders.}  
{\small \it Source: New York Times (January 2002)}.
 \end{figure}

FMR is the world's largest investment fund with assets estimated in 2003
at about 1 trillion dollars. The parent company has several subsidiaries
such as Fidelity Magellan, Fidelity Growth Company, Fidelity Leveraged
Company which manage its funds. Unfortunately, in contrast to the
previous case, we do not know precisely the reason of Fidelity's move.
Of course, if Kmart had been unable to avoid bankruptcy that would have
been a sufficient motive for (as we explain below) share holders usually
lose next to 100\% of their assets in a bankruptcy. However, my reading
is that Kmart was not driven into bankruptcy by its debt but rather by
the withdrawal of its major share holders. That feeling relies on the
fact that Kmart's assets in terms of real estate (land and stores) was
estimated at \$ 15 billions in a report published by the Deutsche Bank
in July 2004%
\qfoot{In late July 2004 Kmart actually sold 78 of its 2000 stores
for about \$ 1 billion.}%
; no doubt that, in early 2002 prior to the bankruptcy,
the value of this asset was substantially larger. As Kmart's debt never
exceeded \$ 4 billions, it was not in an Enron-like situation where
the debt was actually larger than the real worth of the assets. 
\qpar

Naturally, if major investors withdraw their support, if the company's debt
is downgraded by rating agencies (which indeed happened on January 17, 
2002), then it cannot get new short-term loans and bankruptcy becomes
unavoidable. At this point there is a question for which we do not
yet have a satisfactory answer. We know that Fidelity (along with other
mutual funds) sold its shares, but we do not know who bought them.
In a climate where there are persistent rumors about a possible 
bankruptcy there must have been very little incentive for any investor
to buy these shares even at a fraction of their initial price. 
Of course, on the NYSE it is the duty of market makers to ensure
market fluidity, but it is difficult to understand how can they fulfill this
task when there are no buyers whatsoever.
\qpar

Before we turn to the next episode we need to better understand what
happens once a corporation has filed for chapter 11 bankruptcy protection,
especially with respect to its shares and shareholders.

\qI{Impact of bankruptcy on stocks}

After the bankruptcy, some of the major debt holders usually provide
short-term cash to the bankrupt company; in return, they get highest
priority on the list of the creditors. On the contrary, common share holders
are listed at the bottom. However, it should be noted that a company is
not automatically delisted from the exchange after asking for 
bankruptcy protection. The only immediate change is the fact
that the ticker symbol becomes followed by the letter Q. 
Thus Kmart's ticker symbol was
changed from KM to KMQ. Incidentally, one week after the bankruptcy
the price stood at \$ 1.4, more than double the \$ 0.66 the stock
was worth the day the bankruptcy 
was declared; but this improvement did not last very long.
On the New York Stock Exchange the criteria
for continued listing include a requirement that a company's stock trade
at a minimum average price of 1 dollar over a 30-day period. 
In the case of Kmart, the share traded under the \$ 1 threshold from
July to December 2002 and, as a result, it was delisted from the NYSE
in December. But even after that, the stock continued to be traded
in Pink Streets, an over the counter exchange. Yet, when Kmart
emerged from bankruptcy on May 6, 2003, it canceled its old stock and issued
new shares. At this point the old share holders
lost all their remaining assets. Half of the stock issued on May 6, went to
creditor Edward Lampert. The rest of the shares went to other creditors.
With 49\% of the shares, Lampert got complete control over Kmart. He and
some of his associates, including William Crowley, soon after became part
of the board of directors comprising nine people (Detroit Free Press,
May 6, 2003).
Thus began a new phase
of Kmart's history to which we turn now. 

\qI{Lampert's era}

Although still a fairly young man (he is born in 1963) Edward Lampert
was in 2003 one of the stars of the hedge fund industry; he was
listed in fourth position among the top 10 fund managers with an
annual pay of
\$ 420 million (Georges Soros came first on this list with \$ 750 million).
Fig. 2d shows a strong and steady price increase after Kmart
emerged from bankruptcy. 
  \begin{figure}[htb]
    \centerline{\psfig{width=12cm,figure=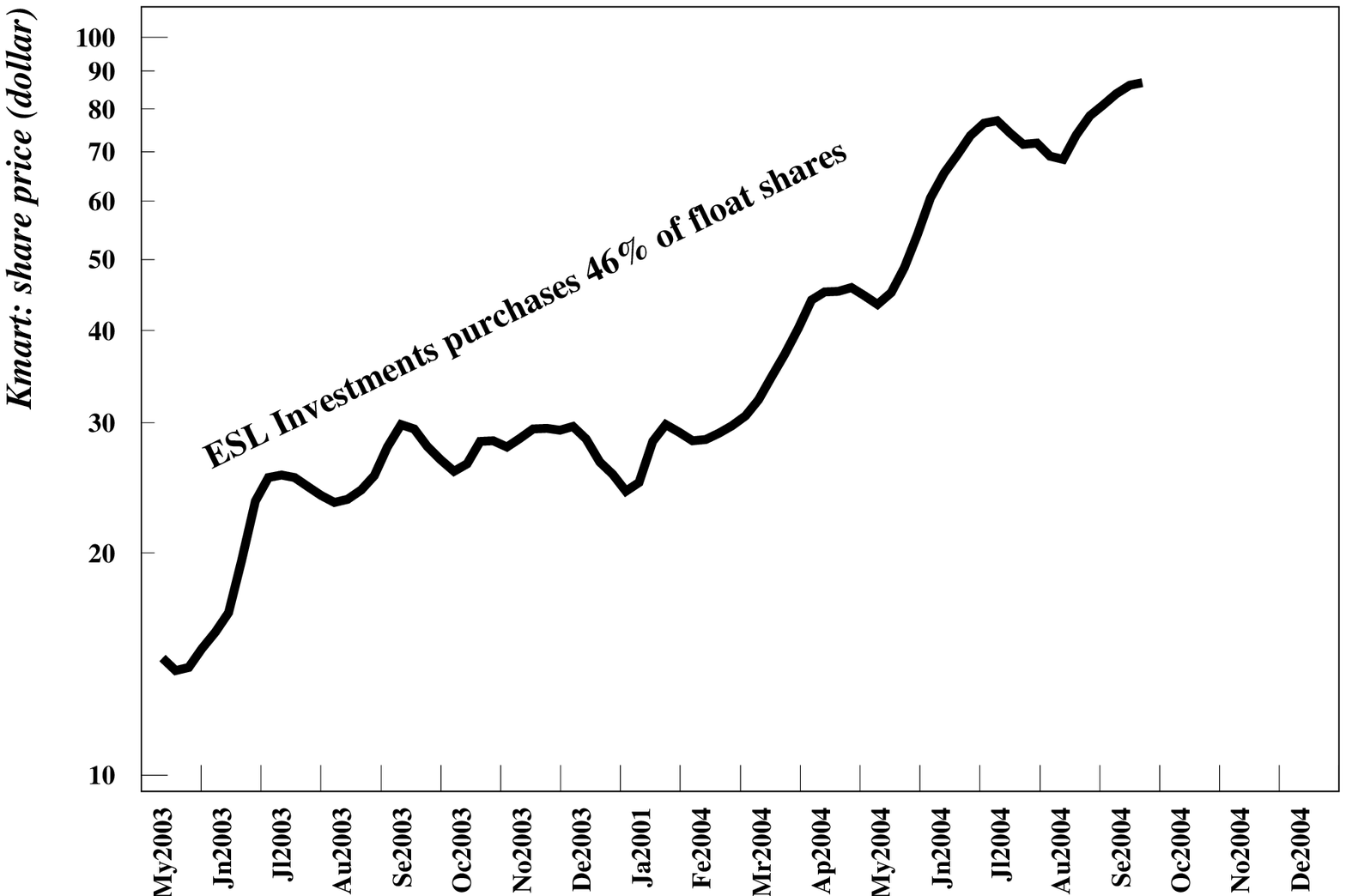}}
    {\bf Fig.2d Kmart share price (May 2003 - September 2004)}
{\small After Kmart emerged from bankruptcy, 49\% of its shares were
in the hands of ESL Investments, one of its creditors. SEC reports show 
that over 2003-2004 ESL bought a substantial slice of the remaining float.}  
{\small \it Source: http://finance.yahoo.com, table 1}.
 \end{figure}

Does this mean that Kmart had solved
its problems and was up for a new start?
Certainly not. Kmart's market share continued to shrink and in fact
at a faster rate than before the bankruptcy. In 2003, comparable-store
sales dropped by 30\% and, even more worrying, the rate accelerated
from 3.2\% in the first quarter to 13\% in the last quarter. Yet, over that
time interval the share price more than doubled. How can one explain
that?
\qpar

The answer is very simple. There was a permanent flow of purchases
by Lampert's hedge fund (see table 1). It should be noted that all
these transactions were performed in the Non-Open market. In this
market, the price is settled by a prior agreement between buyers and
sellers. As all the shares were in the hands of various institutions, it is
very likely that strategic considerations played a big role in these transactions.
The financial situation of Kmart did indeed improve but only because
it sold some of this stores to Sears, Roebuck and Co. and to Home Depot%
\qfoot{Incidentally, it can be noted that Lampert also owned 28\% of 
Sears, Roebuck and Co.}%
.



\begin{table}[htb]

 \small 

\centerline{\bf Table 1\quad Purchases of Kmart shares by Lampert's 
hedge fund in 2003-2004}

\vskip 3mm
\hrule
\vskip 0.5mm
\hrule
\vskip 2mm

$$ \matrix{
\tvi 
& \hbox{Year}  & \hbox{Date} 
& \hbox{Number of shares}  \cr
\qtb  & \hbox{}  & &\hbox{[million shares]} \cr
\noalign{\hrule}
\qth 
1&2003 & \hbox{June 30} &  7.0\cr
2&2003 & \hbox{Oct. 23} &  1.7\cr
3&2003 & \hbox{Nov. 03} &  0.45\cr
4&2003 & \hbox{Feb. 12} &  0.024\cr
5&2004 & \hbox{Apr. 27} &  0.89\cr
6&2004 & \hbox{Jul. 1} &  7.2\cr
7&2004 & \hbox{Jul. 16} &  2.1\cr
8&2004 & \hbox{Aug. 18} &  0.053\cr
& & \hbox{} &  \cr
&\hbox{\bf Total} \hfill && \hbox{\bf 19.4} \cr
 &\hbox{\bf \% of shares} \hfill && \hbox{\bf 23\%} \cr
\qtb &\hbox{\bf \% of float} \hfill && \hbox{\bf 46\%} \cr
\noalign{\hrule}
} $$

\vskip 1.5mm
Notes: As of August 24, 2004, Lampert's hedge fund,
ESL Investments, owned directly of indirectly, 82\% of Kmart
shares, a stake consisting of its initial stake of 49\% as a creditor of
Kmart plus the above 23\% which it purchased over 2003 and
2004. All the acquisitions listed in the table were made in
the Non-Open-Market. 
\qL
Source: Insider and Form 144 Filings - ESL Investments - 
http://biz.yahoo.com/t/97/342.html
\vskip 2mm

\hrule
\vskip 0.5mm
\hrule

\normalsize

\end{table}


\qI{Hints about the future of Kmart}

As the price rise documented in Fig. 2d was largely disconnected from
underlying fundamentals, it can hardly be expected that it will continue
for long. In fact, it will continue until Lampert decides that his strategy
no longer requires the price to rise. Several analysts expect that Lampert
will continue to sell the most valuable assets of Kmart before 
eventually taking it completely out of the retail business. Whether that
can be done in the present framework or requires the liquidation
of the company is a matter of debate. As one analyst lucidly commented,
now that Kmart is under the control of ESL, any 
analysis based on Kmart's fundamentals becomes irrelevant
because what is good for Kmart is not necessarily good for ESL and
vice versa. As of September 4, 2004 about 25\% of the shares were
sold short%
\qfoot{This percentage is well above standard short percentages: for 
instance it was equal to 0.30\% for General Electric and 0.82\% for IBM}%
, a transaction which generates a profit only if
the price falls.
\qpar

{\bf Remark} With over 80\% of the shares in the hands of 
ESL Investments one would expect the trading volume for Kmart's shares
to be markedly lower than for other corporations whose ownership is less
concentrated. Yet, one observes exactly the opposite. Between
July and September 2004, an average 2.7 million shares were traded
daily which represents 3\% of the shares outstanding; that figure is
about 150 times higher than for General Electric and 11 times higher
than for IBM. Why is Kmart trading volume one or two orders
of magnitude higher than such widely traded stocks as GE or IBM? This
remains an open question.

\qI{Summary of Kmart's case}

Let us summarize what we learned from this case-study.
\qee{1)} Until 1999-2000 there was a connection between
Kmart's share price and its achievements as a discount retailer.
\qee{2)} After October 2000, there is a one-year episode marked by 
a strong price rise due to a deal with a supplier which bears no
relationship whatsoever with Kmart's performances.
\qee{3)} The bankruptcy occurred when one of the major share holders
withdrew its support. Although it is difficult to distinguish 
with certainty between cause
and consequence, the question must be examined
in the light of what happened subsequently, namely the fact that the
corporation fell under the control of Lampert's hedge fund. 
\qee{4)} The 700\% price increase between May 2003 and September 2004
was completely at variance with the evolution of
Kmart's growth fundamentals.
\qpar

Kmart was selected because it went through a 
bankruptcy. The idea was that a major shock would reveal features
about the behavior of share holders which are relatively obscured
and hidden in more ordinary conditions. However, similar 
mechanisms are at work also in cases characterized by big
shocks even in the absence of a bankruptcy. This is illustrated by the
following example.

\qI{Converium} 
Converium (NYSE: CHR) is a Swiss reinsurer which ranks among the
top 10 reinsurers and employs approximatively 850 people in 23
countries around the world. Why did I select Converium among many
other possible cases? My attention was attracted to it because it
experienced a sharp price fall in July 2004. Subsequently I discovered
that one of our colleagues, econophysicist Michel Dacorogna, is a 
senior member of its Risk Modeling team; naturally, this further increased
my interest in the company. The graph (Fig. 3) of its share price
is particularly striking because it has been very stable during two years
before dropping sharply by 50\% on July 21, 2004. 
  \begin{figure}[htb]
    \centerline{\psfig{width=12cm,figure=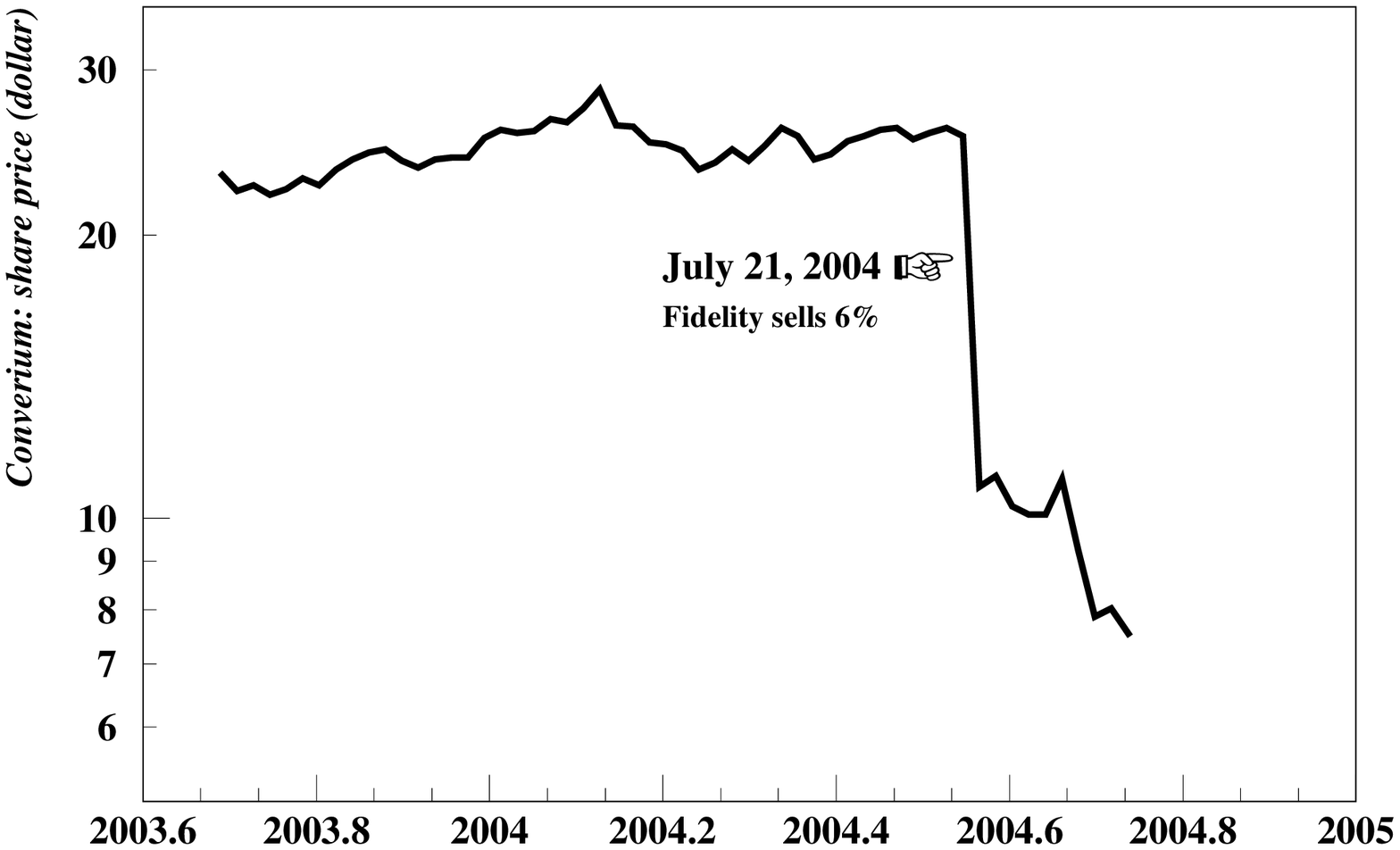}}
    {\bf Fig.3 Converium share price (September 2003 - September 2004.}
{\small Converium, a Swiss reinsurer began to be listed on the 
New York Stock Exchange in January 2002. From that date to mid-July 2004
its price remained within a fairly narrow margin of $ 25 \pm 5 $ dollars.
Then, on July 21, it suddenly dropped 50\% after the company's
announcement that it will have to increase its reserves.}  
{\small \it Source: http:// finance.yahoo.com}.
 \end{figure}
After this
date it continued to fall albeit more slowly. As of September 23, 2004
its share price was as low as \$ 7.80 which means that it had been
divided by more than 4 with respect to the price level of January 2004.
The comments offered by analysts in the wake of the fall of July 21 
were not very convincing. They attributed the fall to a net loss 
amounting to
22\% of its capitalization and to the fact that it
had to strengthen its reserves by
a similar amount. Although fairly serious, such a problem did not
imperil the existence of the company especially because the loss
was limited to its activity in the United States%
\qfoot{As one of the main activities of Converium was the
reassurance of airspace industries, it is possible that the loss was
a consequence of 9/11}%
. 
A more tangible explanation came two weeks later, on August 3 2004,
in the form of the following statement made by the company:
\qdec{Converium Holding hereby informs that Fidelity International
(based in Hamilton, Bermuda) has reduced its holding in Converium 
from 9.87\% to 3.81\%.}
\qpar

As is common in such announcements, it did not say when exactly Fidelity
had sold its shares%
\qfoot{The reaction of the market to the information revealed on
July 21 was particularly swift, but this was largely due
to the fact that the ``market'' was a small group of fund managers, 
among whom the
manager of Fidelity International had a leading role.}
.
As in the case of Kmart in January 2002, a flow of bad news followed.
For instance, on September 1, Standard and Poor's
cut the rating of Converium's North America to BBB, just one notch above
junk status. The rating
of Converium AG, the Swiss unit was also lowered. In this respect, it should
be noted that in principle the role of rating agencies is to foresee possible
financial problems ahead of the ``market'', whereas in this case 
as well as in many others (e.g. Enron, Kmart, WorlCom)
their reactions {\it followed} the 
announcement made by the company, in the present
case  by more than one month. 
\qpar

At first sight, one may be tempted to think that these reactions were 
simply the normal consequence of a change in the growth 
fundamentals of the company.
However, the fact that on August 10, Capital Group, another mutual fund, 
{\it increased} its stake in Converium from 4.05\% to 5.34\% shows that
the fundamentals were not read in the same way by all the players. 
In fact, it seems it was rather a showdown between two groups of
players.
\qpar

On September 15, 2004 came the first rumors
that Converium could become the
target of a possible takeover, an operation that its low stock price 
facilitated. The most widely named potential buyers
were Munich Re and Berkshire Hathaway, Warren Buffet's company.
On September 17, Converium made known that its discussions to
enter into a possible partnership with these two companies were
nearing a successful conclusion.

\qI{Generalizations}

To what extent is it possible to generalize the results of these case-studies?
One can give the following answers.
\qbu The Kmart episode was not an isolated example. As a matter of fact,
the strategy Lampert used at Kmart had been used previously 
in others of its acquisitions such as Autonation (NYSE:AN), America's
largest retailer of new and used vehicles,  Autozone (NYSE: AZO), 
Deluxe (NYSE: DLX) and finally Sears, Roebuck and Co (NYSE: S).
\qbu The fact that an investment fund reduces its stake in a company
to a considerable extent in a short time interval is relatively common.
Table 2 gives a number of illustrations over 2002-2004 for FMR which
naturally is only one of the giant mutual funds (albeit the largest).
Usually the growth fundamentals of a company
do not change sharply in a few months which means that such massive
sales (or purchases) pursued broader strategic objectives. 



\begin{table}[htb]

 \small 

\centerline{\bf Table 2\quad Effect on share prices of a change in
holdings by Fidelity Management and Research (FMR)}

\vskip 3mm
\hrule
\vskip 0.5mm
\hrule
\vskip 2mm

$$ \matrix{
\tvi 
&\hbox{Date}  \hfill & \hbox{Company} \hfill& \hbox{Ticker} \hfill & 
\hbox{Initial} & \hbox{Subsequent} & \hbox{Price} \cr
 &\hbox{}  \hfill & \hbox{} \hfill& \hbox{symbol} \hfill & 
\hbox{stake} & \hbox{stake} & \hbox{variation} \cr
\qtb
 &\hbox{}  \hfill & \hbox{} \hfill& \hbox{} & 
[\%] & [\%] & [\%] \cr
\noalign{\hrule}
\qth 
 1&\hbox{2002 Dec.}  \hfill & \hbox{Teradyne} \hfill& \hbox{NYSE:TER} \hfill & 
15 & 8.24 & -61 \cr
2 &\hbox{2003 Dec.}  \hfill & \hbox{Delta Airlines} \hfill& \hbox{NYSE:DAL} \hfill & 
 7.3&  0.5& -72 \cr
3 &\hbox{2003 Oct.}  \hfill & \hbox{Forrester Research} \hfill& 
\hbox{NASDAQ:NM} \hfill & 
9.4 & 2.8 & -30 \cr
\qtb
4 &\hbox{2004 Feb.}  \hfill & \hbox{Boeing} \hfill& \hbox{NYSE:BA} \hfill & 
 2.2& 3.6 & \phantom{-}17 \cr
\noalign{\hrule}
} $$

\vskip 1.5mm
Notes: FMR is the world's largest investment fund with about one trillion
dollar under management (which represents 10\% of the US GDP). 
The price variation refers to the quarter during which the sales or purchases
were made (we do not know the exact dates of the transactions). 
Earlier FMR moves include the reduction of its stake in
(i) United Airlines from 6\% to 2\% (June 1994),
(ii) Apple Computer from 11\% to 2.5\% (August 1995),
(iii) Technology stocks (end of 1995),
(iv) US Airways from 11.3\%  to 5.8\% (May 1996),
(v) Digital Equipment Corporation from 13.7\% to 7\% (June 1996),
(vi) Chrysler Corporation from 12.2\% to 7.8\% (June 1996)
Besides
FMR there are several other mutual funds giants (e.g. Vanguard Group, 
Capital Research and Management, State Street) whose moves
have also a substantial impact on stock prices.
\qL
Sources: Boston Business Journal (Dec. 10 2002, Oct. 10 2003);
Atlanta Business Chronicle (Dec. 19 2003); The News Tribune of Tacoma,
Washington (Feb. 18 2004), New York Times (June 11 1994, January 12 1996,
Aug. 15 1996);
Wall Street Journal (Oct. 12 1995); USA Today (May 9 1996);
Boston Herald (July 11 1996).
\vskip 2mm

\hrule
\vskip 0.5mm
\hrule

\normalsize

\end{table}


\qI{Conclusion}

The main message of this paper is the observation that many of
the major shocks to which
companies are confronted are due to the moves of a small number
of investment funds. In the case of Kmart we have seen that
single investors played a central role in each of 
the three successive episodes which sealed the fate of the corporation
between 2000 and 2004:
first it was Burkle, then FMR and finally Lampert.
\qpar

The key role played by major investment funds can be further
illustrated by comparing the major holders of three airline companies, 
namely American Airlines, Delta Airlines, and US Airways. 
Table 3 summarizes the information as of September 10, 2004.



\begin{table}[htb]

 \small 

\centerline{\bf Table 3\quad Major holders in three airlines}

\vskip 3mm
\hrule
\vskip 0.5mm
\hrule
\vskip 2mm

$$ \matrix{
\tvi 
&\hbox{Holder}  \hfill & \hbox{Stake in} & \hbox{Stake in} & \hbox{Stake in} \cr
  & & \hbox{American Air.} &  \hbox{Delta Air.} &  \hbox{US Air.} \cr
\qtb &   &  \hbox{[\%]} &\hbox{[\%]} &\hbox{[\%]} \cr
\noalign{\hrule}
\qth 
\hbox{\bf A} \hfill & \hbox{\bf Insiders and rule 144 holders} \hfill & <1 & <1 & \cr
 & \hbox{Retirement System of Alabama} \hfill & & & 79 \cr
 & & & & \cr
 \hbox{\bf B} \hfill & \hbox{\bf Institutions} \hfill &  &  & \cr
 & \hbox{PAR Capital Management} \hfill & 9.6& & \cr
 & \hbox{Prime Cap Management} \hfill & 8.8 & 10 & \cr
 & \hbox{Lord Abbet} \hfill & 5.4 & 4.8& \cr
 & \hbox{Brandes Investment Partners} \hfill & 3.3 & 9.4 & \cr
 & \hbox{Wellington Management} \hfill & 4.9 & & \cr
 & \hbox{Stavo Asset Management} \hfill & 4.5 & & \cr
 & \hbox{Barclays Bank} \hfill & 4.0 & & \cr
 & \hbox{Capital Guardian Trust} \hfill & & 9.1& \cr
 & \hbox{Capital Research and Management} \hfill & & 5.5& \cr
 & \hbox{UBS Global Asset} \hfill & & 3.9& \cr
 & \hbox{State Street} \hfill & & 3.2& \cr
 & \hbox{Baupost Group} \hfill & & & 4.3\cr
 & \hbox{Farallon Capital Management} \hfill & & & 1.6\cr
 & \hbox{} \hfill & & & \cr
 \hbox{\bf C} \hfill& \hbox{\bf Funds} \hfill & & & <0.1\cr
 & \hbox{Vanguard/Primecap} \hfill & 5.8& 6.5& \cr
 & \hbox{Lord Abbett Fund} \hfill & 2.2& 2.7& \cr
 & \hbox{Vanguard Horizon} \hfill & 1.9 & 2.3 & \cr
 & \hbox{Fidelity Growth} \hfill & 9.9 & & \cr
 & \hbox{Vanguard Windsor Fund} \hfill & 2.0 & & \cr
 & \hbox{} \hfill & & & \cr
\qtb
 & \hbox{\bf Percentage held by institutional owners} \hfill & 
\hbox{\bf 97}& \hbox{\bf 97} & \hbox{\bf 88} \cr
\noalign{\hrule}
} $$

\vskip 1.5mm
Notes: 
Although it may not be of great relevance, especially
in major shocks,
we retained the standard distinction between insiders 
and rule 144 holders (A), institutions
(B), and funds (C). Most of the institutions which have substantial
stakes in both American and Delta also hold shares of other airlines such
as Continental, NorthWest or SouthWest. The table documents the
sharp difference between the ownership structure of American and
Delta on the one hand and US Air on the other hand. The giant mutual
funds have no longer any substantial stake in US Air.
\qL
Source: http://finance.yahoo.com/
\vskip 2mm

\hrule
\vskip 0.5mm
\hrule

\normalsize

\end{table}


Three significant observations can be made.
\qee{1)} Several institutional holders have a stake in both American
and Delta. Examination of other airlines 
(e.g. Continental, NorthWest, Southwest)
shows that these players
also hold substantial stakes in those other airlines.
\qee{2)} There is a fundamental difference between the major holders
of American and Delta on the one hand and those of US Airways on the
other. In the latter we do not find any major investment funds with
a substantial (say over 1\%) stake.
Most of the shares are in the hands of the
Alabama Retirement Fund which, through its links with Social 
Security, is probably partly funded by federal money. This striking difference
is certainly to be attributed to the fact that US Airways went through 
bankruptcy in August 2002. As seen previously, stocks are likely to
lose all their worth in a bankruptcy process. For major holders the main
problem therefore is to be able to sell before the price
has collapsed. Naturally, such tactics are double edged
because the
withdrawal of a major holder may drive down the market price
to a point which makes bankruptcy ineluctable.
\qee{3)} On financial websites such as Yahoo, investment companies
are listed apart  from the funds itself. One may wonder if such a distinction
is really relevant. Consider for instance the Vanguard Group which
offers more than 100 funds. It can be admitted that in ordinary day-to-day
operations, each fund has some autonomy. However, in critical
junctures (such as a bankruptcy risk) all funds tend to follow the same
tactic as can be seen from the fact that {\it all} Vanguard funds 
left  US Airways as it was stumbling toward its first bankruptcy%
\qfoot{Furthermore different investment companies may have overlapping
interest; as an example one can mention that Wellington Management, one
of the oldest American money management firm, manages 16 of the 
funds offered by Vanguard.}%
.
\qpar

Fig. 4 shows that, since 1945, mutual funds experienced an exponential
growth which was shortly interrupted only by the bear market of 1968-1978.
  \begin{figure}[htb]
    \centerline{\psfig{width=12cm,figure=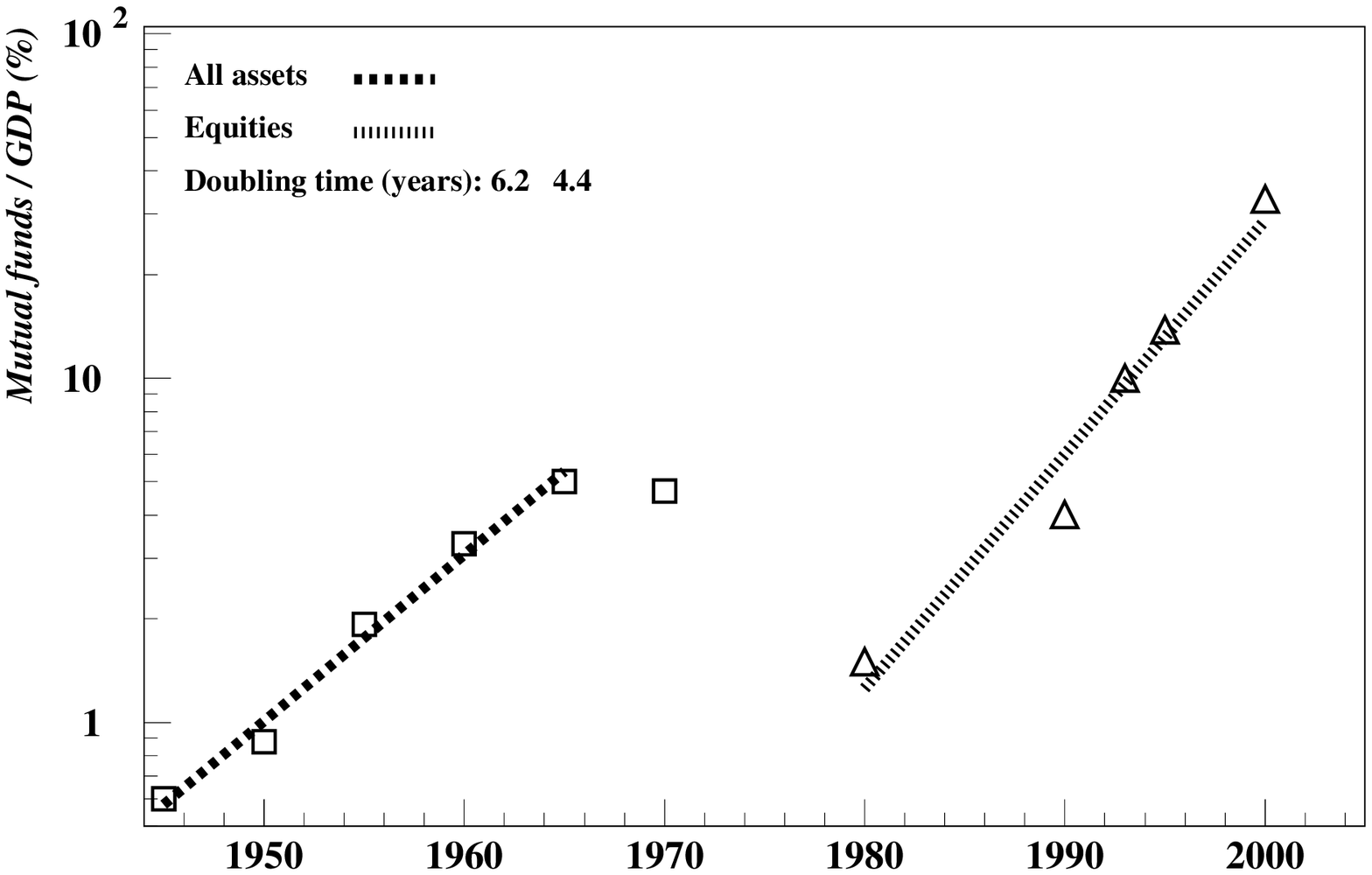}}
    {\bf Fig.4 Growth of American mutual funds compared to the growth of the
US gross domestic product (1945-2000).}
{\small Between 1947 and 2000 the assets of mutual funds as a proportion
of GDP have been multiplied by a factor of the order of 100.
We had to rely on two different series because the pre-1970
data available in the Historical Statistics of the United States refer
to total assets (i.e. stocks plus bonds), whereas the Statistical Abstract
more specifically gives equity assets. Both trajectories are exponential.
Naturally, mutual funds represent only one class (albeit the most important)
of institutional share holders besides insurance companies, banks, 
state retirement funds, hedge funds, etc.}  
{\small \it Source: Historical Statistics of the US (1975), Statistical Abstract
of the US (various years)}.
 \end{figure}
As for any exponential growth, the beginnings were inconspicuous. 
It is only in recent years that mutual funds were able to get a firm
grip on American stock markets. If this evolution continues the
conception based on micro-players will become
less and less relevant. 
\qpar

In a previous paper (Roehner 2005) it was shown that,
through buyback programs,
corporations can influence the price level of their
own stock. To make that point we
did not have to resort to microeconomic analysis as we did
in the present paper; why did the present study require behavior analysis
at the level of individual players?
The answer is obvious. We wanted to scrutinize the economic rationale
of the moves of major players and in order to do that one has
to understand and weigh their actions in detail.
Such an approach is complimentary to the comprehensive
macrodynamic analysis of market structure carried out
by other researchers such as for instance Elroy
Dimson et al. (2002), Rosario Mantegna et al. (2000) or
Didier Sornette (2003). Finally,
there is an important question which we did not consider
and which should be addressed in a subsequent study. 
What is the kind of interaction between macro-players. Is it a competitive or
cooperative linkage, or perhaps both depending on circumstances?
\qpar

{\bf Acknowledgements}\quad I am most grateful to Stanislaw Drozdz,
Olivier G\'erard, Cornelia K\"uffner and Joseph MacCauley 
for their helpful comments.

\vfill \eject

 \appendix

 \qI{Appendix A: Sources and secretiveness}

In many respects this study would not have been possible without the
Internet. In pre-Internet times, one would have had to rely on 
archives of newspapers and economic magazines. 
Because of the impossibility of
performing broad keyword searches, many relevant papers would not
have been identified. In addition, locating and accessing
the relevant archives in various
research libraries would have made the whole process
utterly time-consuming. 
Apart from the Internet we also used Lexis-Nexis, 
a newspaper data base which is available to subscribers (many research
libraries offer this service).
However, even with these tools, it was not always possible to get all
the information that would have been needed. 
There were two main obstacles.
(i) Some information is not made public. For instance, hedge funds
are not required to report their positions, trading activity and
creditworthiness. By 2000, there were about 5,800 hedge funds,
most of them registered off-shore to avoid taxation, totaling \$\~300 billion
in capital (Derivatives Study Center, http://www.financialpolicy.org).
As an illustration of this secretiveness, one can mention
the episode of the speculation against 
the British pound in 1992 which led to its
withdrawal from the European Monetary System. This destabilization
is commonly attributed to Georges Soros' hedge fund, but due to the 
lack of any official statement it is still impossible to confirm or to
disprove this assertion.
\qL
ii) A second problem concerns the fact that even when it is made 
public the relevant information often comes too late. The major source
of information about ownership are the reports required by the
Securities and Exchange Commission in which companies document
their most recent moves. Unfortunately, these reports are published
only every quarter (in some cases every semester) which means that
the information will come weeks after the move has affected price
levels. For instance, in the case of Kmart we learned about the
the withdrawal of Fidelity Management by a report that came on
February 2, 2002 that is to say almost one month after the price
collapse began and two weeks after Kmart filed for bankruptcy 
protection. 
\qpar

There is a last point which has to be mentioned. So far we did not
consider the link between mutual funds and their subscribers. One could
argue that subscribers can influence the behavior of fund managers.
This would be true if they could get informed in time. As we have seen,
this is not the case: not only do they learn about major moves well after
the event, but usually they are also unable to know how fund managers
who sit as directors on corporate boards voted in crucial occasions
\qfoot{In January 2003, in spite of vocal opposition led by Fidelity Investments
and Vanguard,
the Securities and Exchange Commission, voted
in favor of disclosure of proxy votes by mutual funds and established August 31,
2004 as the deadline for disclosure of votes cast during the year ending
June 30, 2004 (see ``Behind the curtain'' by the AFL-CIO Office of investement,
September 2004). We observe once again that
this decision makes information available to shareholders
only months after the events took place.}
.

\vfill \eject

{\large \bf References}

\vskip 5mm

\qparr
Common stock price histories 1910-1987. Logarithmic supplement (1988)
WIT Financial Publishers. Morristown (New Jersey).

\qparr
Dimson (E.), Marsh (P.), Staunton (M.) 2002: Triumph of the
optimists: 101 years of global investment returns. 
Princeton University Press, Princeton.

\qparr
Kotz (D.M.) 1978: Bank control of large corporations in the
United States. University of California Press. Berkeley. 

\qparr
Mantegna (R.N.), Stanley (H.E.) 2000: An introduction to econophysics:
correlations and complexity in finance. 
Cambridge University Press, Cambridge.

\qparr
Roehner (B.M.) 2005: Stock markets are not what we think they are:
the key roles of cross-ownership and corporate treasury stock.
Physica A 347, 613-626.

\qparr
Sornette (D.) 2003: Why stock markets crash: critical events in
complex financial systems. 
Princeton University Press, Princeton.

\end{document}